\documentclass [a4paper,10pt]{article} 
\usepackage[ascii]{inputenc}
\usepackage{mathtools}
\usepackage{url}
\usepackage{amsfonts}

\title{Do symmetries ``explain'' conservation laws? The modern converse Noether theorem \emph{vs} pragmatism.}

\author{ 
Harvey R. Brown\\
Faculty of Philosophy, University of Oxford,\\ Radcliffe Observatory Quarter 555,  
Woodstock Road,\\ Oxford OX2 6GG, U.K.}

\begin{document}

\maketitle

\bigskip

\begin{abstract}

Noether's first theorem, in its modern form, does not establish a one-way explanatory arrow from symmetries to conservation laws, but such an arrow is widely assumed in discussions of the theorem in the physics and philosophy literature. It is argued here that there are pragmatic reasons for privileging symmetries, even if they do not strictly justify explanatory priority. To this end, some practical factors are adduced as to why Noether's direct theorem seems to be more well-known and exploited than its converse, with special attention being given to the sometimes overlooked nature of Noether's converse result and to its strengthened version due to Luis Mart\'{i}nez Alonso in 1979 and independently Peter Olver in 1986.
\end{abstract}

\tableofcontents{}

\section{Introduction} 

In 2004, I argued with Peter Holland \cite{BH} that in her celebrated 1918 paper, Emmy Noether established by way of her (first) theorem\footnote{For an outline of a special case of the first theorem, see section 4.1 below. English translations of Noether's paper \cite{Noether} by M. A. Tavel and Y. Kosmann-Schwarzbach are found in \cite{Tavel} and \cite{K-S1}, pp. 1-22, respectively. For an account of Noether's second theorem see, e.g., \cite{BB}.} a \emph{correlation} between variational symmetries and conservations laws, but not an explanatory one-way arrow from the former to the latter. Our argument was based on the fact that Noether also provided a converse of the theorem in her paper. But as we shall see below, this converse result is not sufficient to make the point: appeal must instead be made to the modern, generalised version of the joint direct and converse theorem due to Luis Martinez Alonso in 1979 and independently Peter Olver in 1986. A careful scrutiny of the conditions of the theorem will be given below.

Despite this strengthening of the argument, such an egalitarian position in the reading of Noether's theorem is hardly in tune with 
typical discussions of the theorem in the literature: conservation principles are usually taken to follow from the existence of variational symmetries. Attempts have been made to establish the priority of symmetries on metaphysical grounds.\footnote{Mark Lange, for instance, has provided an argument to the effect that symmetry principles possess ``a variety of necessity'' that is stronger than that of conservation laws \cite{Lange}; note that Lange does not base this claim on Noether's theorem \emph{per se}. A hard-hitting critique of Lange's position is found in  \cite{Smith}. A further worry is mentioned in section 5 below.} However, in this paper, a more pragmatic approach is taken, looking at how symmetry principles are actually used by physicists within the Lagrangian framework. This may not strictly establish an explanatory fundamentality for symmetries in relation to conservation principles, but it may help to explain why more emphasis is placed in practice on Noether's direct theorem than on its modern converse.

\section{The standard account}

The following quotations seem to typify discussions of the relationship between symmetries and conservation principles in the literature. We start with Landau and Lifshitz in 1960:
\begin{quote}
There are some [integrals of the motion] whose constancy is of profound significance, deriving from the fundamental homogeneity and isotropy of space and time ... \footnote{\cite{LL}, section 6.}
\end{quote}
In his 1986 book \emph{Fearful Symmetry}, the theorist and expositor A. Zee wrote:
\begin{quote}
For years, I did not question where these [energy, linear and angular momentum] conservation laws came from; they seemed so basic that they demanded no explanation. Then, I heard about Noether's insight and I was profoundly impressed. The revelation that these basic conservation laws follow from the assumption that physics is the same yesterday, today, and tomorrow; here, there, and everywhere; east, west, north, and south, was for me, as Einstein put it, essentially spiritual.\footnote{\cite{Zee1}, p. 121. This statement is striking because Zee also writes that ``Noether's observation tells them [physicists aware of a conserved quantity] that the action must have a corresponding symmetry.'', p. 121.}
\end{quote}
Lewis Ryder, in his 1996 book \emph{Quantum Field Theory}, referred to ``the origin of the conservation laws for energy, momentum and angular momentum; they all follow from space-time symmetries via Noether's theorem \ldots.''\footnote{\cite{Ryder}, section 3.2.}
In 2003, an article \cite{HTH} in the \emph{American Journal of Physics} claimed that
\begin{quote}
When the German mathematician Emmy Noether proved
her theorem, she uncovered the fundamental justification
for conservation laws. 
\end{quote}
In their 2004 book \emph{Classical Mechanics}, Kibble and Berkshire stated that the ``physical reasons" for the existence of conserved quantities are
``expressions of symmetry properties expressed by the system''\footnote{\cite{KB}, p. 291.} and that ``any symmetry property of the system leads to a corresponding conservation law \ldots''\footnote{\emph{Op. cit.}, p. 301.}

 Two prominent voices on symmetries in the philosophical literature are Bas van Fraassen and Mark Lange. Writing in 1989, van Fraassen asserted:
\begin{quote}
In the twentieth century we have learned to say that every symmetry yields a conservation law.\footnote{\cite{vF}, p. 258.}
\end{quote}
He went on to say that symmetries ``engender'' conservation laws.\footnote{\emph{Op. cit.} p. 433.} More recently Lange has referred approvingly to the ``commonly held view'' that ``a given symmetry principle explains the associated conservation law''.\footnote{\cite{Lange}, p. 462.}

Note that neither Landau and Lifshitz nor Kibble and Berkshire mention Noether's theorem explicitly, and it must not be thought that demonstrating the connection between symmetries and conservation laws, in either classical or quantum mechanics, necessarily requires appeal to Lagrangian methods.\footnote{Nor should it be overlooked that (i) the existence of conservation laws, or integrals of the motion, do not depend in general on the existence of symmetries and (ii) invariance with respect to part or all of the Galilean group, for example, does not invariably imply in Newtonian mechanics the existence of any of the standard conservation laws. For further details see \cite{Havas},  pp. 148-150.}At any rate, the supposed explanatory priority of symmetries reflected in all of the remarks cannot be regarded as obvious, for several reasons. First, the real meat in the physics ultimately resides in the equations of motion or field equations (Euler-Lagrange equations in the case of Lagrangian systems) which are the source of \emph{both} the dynamical symmetries and the conservation laws. From this point of view, it is hard to see any distinction between them of the kind we are interested in. Second, in relation to Noether's theorem, there is in general no unique symmetry associated with a given conservation law, or \emph{vice versa}, as we shall see below. Third, even if certain field equations were, as a matter of historical fact, discovered by imposing certain symmetries or conservation laws as \emph{a priori} constraints, this doesn't settle the fundamental explanatory issue. (We return to this point in section 6.2.) And fourth, as mentioned above, Noether's theorem has a converse, which in its modern form arguably deserves to be better known. 
But before turning to the converse theorem, it is worth reminding ourselves of the subtleties and limitations of Noether's direct theorem which are not always recognised in conventional statements of its content.

\section{Subtleties of Noether's theorem}

 \begin{enumerate}

\item Not all differential equations in physics have a variational (Lagrangian) formulation; in such cases the application of Noether's theorem is ruled out. Examples are the Navier-Stokes equations (in most cases) and Fourier's heat equation.\footnote{The heat equation features in section 4.3 below. In Appendix A it is shown how a ``trick'' allows for a variational formulation. This formulation is analogous to that for the complex version of the heat equation: the time dependent Schr\"{o}dinger equation for the free particle.

In 1954, Wigner \cite{Wigner54} gave the example of a non-Lagrangian mechanical system obeying Newton's first law of motion, but whose force law is associated with velocities rather than accelerations. In the classical case there are no conservation laws for energy and angular momentum associated with the time translation and spatial isotropy symmetries.} Furthermore, the very question whether a given dynamical theory has a Lagrangian formulation can depend on how the relevant equations of motion, or field equations, are formulated.\footnote{The standard vectorial formulation of Maxwell's theory in terms of $\mathbf{E}$ and $\mathbf{B}$ fields does not have a Lagrangian formulation, but that in terms of potentials does; see \cite{Olver}, p. 289, exercise 4.6. However, a Lagrangian formulation in terms of $\mathbf{E}$ and $\mathbf{B}$ does exist  for the partial, time-dependent Maxwell equations (see the independent work \cite{And} and \cite{Rosen} as well as \cite{Ibrag}), and the full set of vacuum equations also has an unusual vector Lagrangian formulation in terms of the electromagnetic field tensor $F^{\mu\nu}$ and its dual; see \cite{Sud}.

Another example is the equation for a single particle under the influence of a frictional force: $m\ddot{\mathbf{q}} = - k\dot{\mathbf{q}}$. The equation is symmetric under space and time translations, but has no standard momentum or energy conservation laws. It also has no variational formulation, but an equivalent equation is obtained in the usual fashion from the Lagrangian $(1/2)m\dot{\mathbf{q}}^2 \exp [(k/m)t]$, neither of which is time translation symmetric. So the existence of the conserved energy-like quantity $(1/2)m\dot{\mathbf{q}}^2 \exp [2(k/m)t$ does not follow from Noether's theorem; see \cite{Havas} pp. 152-3.}

\item Noether's theorem does not apply to discrete symmetries of equations of motion. But not every continuous dynamical symmetry is a variational (quasi)symmetry.\footnote{A typical case is that of equations of motion which have a scaling symmetry. See \cite{Olver}, pp. 252, 282, and \cite{BH2}, section IV. Other examples in field theory are found in \cite{Boyer} p. 448. The definition of a quasi-symmetry is a transformation of the independent and/or dependent variables such that the action is invariant up to a surface term, so the Lagrangian is invariant up to a total divergence. A special case is described in section 4.1 below, following equation (3).} Nor is it strictly true that every variational (quasi)symmetry is a dynamical symmetry, though this is usually the case.\footnote{A necessary condition is that every dependent variable is dynamical, i.e. that it is subject to Hamilton's variational principle, or equivalently that it satisfies a non-trivial Euler-Lagrange equation; see \cite{BH}, p. 1138, and \cite{Butt}, p. 56.} 

\item The connection between a given conservation principle and its associated symmetry will in general depend on the choice of Lagrangian. Standardly, conservation of energy, linear and angular momentum, for example, are associated via Noether's theorem with the symmetries of time translation (temporal homogeneity), space translation (spatial homogeneity) and rotation (spatial isotropy), respectively, but these associations are only ``mostly'' valid.
One can find Lagrangian systems that are counterexamples in the forward sense and others that are counterexamples in the reverse sense.\footnote{There are systems in which conservation of energy, for example, is not related by the generalised Noether theorem (see below) to temporal homogeneity, and others in which the opposite holds; and similarly for the other cases above of conservation laws. Instructive examples, not always contrived, are found in \cite{Sant} Chart A.11, pp. 340-344, pp. 348-349, and Example A7, pp. 354-356. See Occurrence 2, p. 341, for a counterexample to Lewis Ryder's claim: ``Conservation of energy and momentum, then, holds for any system whose Lagrangian (and therefore action) does not depend on $x^{\mu}$." (\cite{Ryder}, section 3.2).} Typically such nonstandard relationships in the literature arise because some systems share distinct non-gauge-related Lagrangians, i.e. not related by a total divergence.\footnote{Such is ubiquitous in two-dimensional systems, though this is not a necessary condition.}   One of the most striking examples concerns 
 the free Maxwell field. Lagrangian densities can be found such that the conservation of the energy is related to an \emph{internal} symmetry: duality rotations involving the electric and magnetic field strengths ($\mathbf{E}' = \mathbf{E}\cos \theta - \mathbf{B} \sin \theta; \; \; \mathbf{B}' = \mathbf{B}\cos \theta + \mathbf{E} \sin \theta$)!\footnote{See \cite{Sud} for conservation of the stress energy tensor using a four-component Lagrangian density, and \cite{Ibrag}, section 3, for conservation of energy with a scalar Lagrangian density for the time-dependent Maxwell equations. For further discussion of the role of inequivalent Lagrangians in relation to Noether's first theorem, see \cite{BH}, pp. 1136-1137, \cite{Butt}, p. 56, and \cite{Smith}, section 2.} 
 
However, once the true nature of Noether's theorem is revealed by incorporating quasi-symmetries, it is possible to see the degree to which the notion that, for example, conservation of energy is intrinsically linked to temporal homogeneity is misleading -- without needing to appeal to the occasional existence of systems sharing ``inequivalent'' Lagrangians in the above sense. Hopefully this will be made clear below.
 
 \item Not every non-trivial first integral obtained in applications of Noether's theorem is associated with a conservation principle in the usual sense of the word.
This is particularly evident in non-conservative systems, which contrary to widespread belief, may have a Lagrangian formulation. For instance, in the application of Noether's theorem to the linearly damped oscillator, the first integral associated with time translation symmetry is not energy as normally construed.\footnote{See \cite{Sant}, Occurrence 2, p. 341, and \cite{Smith}, section 4.} In optical systems, the time translation symmetry can lead, by Noether's theorem, to a trivial mathematical identity.\footnote{See \cite{BT} and \cite{Olver}, exercise 5.19 p. 372.} 

\item Noether's theorem normally has to do with temporal evolution, but this is not always the case. We see in section 6.1 below applications in the theory of elastostatics which give rise to `conservation' laws, defined not with respect to time but to spatial coordinates. In general the independent variables in the Lagrangian need not contain a time variable.

\item Finally, in cases where the Euler-Lagrange equations themselves can be expressed as Div $P$ = 0, symmetries can sometimes be found for which the Noether conservation law is precisely this equation.\footnote{For an example in elastostatics, see \cite{Olver}, p. 281, and for examples in quantum mechanics and electrodynamics, see \cite{BH2}, sections III and V. Probably the simplest case is the non-relativistic free particle in one dimension, where in relation to the standard Lagrangian $\mathcal{L} = \frac{1}{2}m \dot{\mathbf{q}}^2$, the conservation law related to spatial homogeneity is $\frac{\mbox{d}}{\mbox{d}t}\left(m\dot{\mathbf{q}}\right) = 0$, which is equivalent to the Euler-Lagrange equation $\ddot{\mathbf{q}} =0$.}

\end{enumerate}

\section{The converse Noether theorem}

\subsection{Noether 1918}

Let us start with what Noether's 1918 converse theorem is not. It is \emph{not} the statement that, given a Lagrangian $\mathcal{L}$, and a conservation law emerging from the associated Euler-Lagrange equations, a strict variational symmetry of the action $\int \mathcal{L}\mbox{d}x$ exists which is associated by the direct theorem with this conservation law.

To get a sense of what Noether proved, let us summarise the punchline of the direct theorem. For the sake of simplicity, we consider an action associated with a system of a single degree of freedom of the form $\int \mathcal{L}\left(q(t), \dot{q}(t), t\right)\mbox{d}t$, $q$ a scalar, where $\dot{q} = \frac{\mbox{d}q}{\mbox{d}t}$. The associated Euler-Lagrange expression $E$ is defined in this case by
\begin{equation}
E \equiv \frac{\partial \mathcal{L}}{\partial q} - \frac{\mbox{d}}{\mbox{d}t}\frac{\partial \mathcal{L}}{\partial \dot{q}}.
\end{equation}
 Now suppose there is a group $\mathcal{G}$ of transformations of $q$ and $t$ that depend analytically on a single parameter $\epsilon$, such that the action is invariant under $\mathcal{G}$.\footnote{This means that $\int \mathcal{L} (q, \dot{q}, t) = \int \mathcal{L} (q', \dot{q}', t)$, where the primes represent transformations under $\mathcal{G}$, $ \dot{q}' \equiv \partial q'/ \partial t'$, and the region of integration on the RHS is the transformation of the arbitrary $t$-interval for the integral on the LHS.} It follows that the action will be invariant under the infinitesimal transformations
\begin{equation}
q' = q + Q\epsilon; \; \; \;  t' = t + T\epsilon,
\end{equation}
where $Q\epsilon$ and $T\epsilon$ are the terms of lowest order in $\epsilon$ in the transformations of $q$ and $t$, respectively\footnote{It is assumed that the parameter $\epsilon$ can be chosen such that $\epsilon =0$ corresponds to the identity transformations. Note that since the range of $t$ in the integral $\int \mathcal{L}\mbox{d}t$ is arbitrary, invariance of the integral implies that the Lagrangian $\mathcal{L}$ is invariant.} -- invariant, that is, up to terms of at least second order in $\epsilon$. If $\epsilon$ does not depend on the independent variable $t$, an exercise in the calculus of variations then leads to the so-called \emph{Noether identity}: 
\begin{equation}
E(Q -\dot{q}T) = \frac{\mbox{d}}{\mbox{d}t}\left[-\frac{\partial \mathcal{L}}{\partial \dot{q}}\left(Q -\dot{q}T\right)  - \mathcal{L}T\right].
\end{equation}
Note that in the case of a quasi-symmetry, which Noether does not consider in her paper, where the first variation $\delta \mathcal{L} = \left(\frac{\mbox{d}}{\mbox{d}t}\mathcal{C}(t, q, \dot{q}, \ldots)\right)\epsilon$, the term $\frac{\mbox{d}}{\mbox{d}t}\mathcal{C}$ must be added to the  RHS of (3).\footnote{See. e.g. \cite{BH} section 2. For a critical analysis of two different approaches in the literature to defining variational quasi-symmetries, see section 3 of \cite{BB}.} Finally, when the symmetry group has $\rho$ parameters, there will be $\rho$ such Noether identities.

This completes the direct, or first,  Noether theorem in the special case of $\mathcal{L} = \mathcal{L}\left(q(t), \dot{q}(t), t\right)$, and it makes no reference to conservation laws.\footnote{Nor does Noether ever use the word ``symmetry'' in her 1918 paper; she is concerned only with invariance of (what is normally called) an action functional, and not explicitly with symmetry properties of the Euler-Lagrange equations.}
Of course Noether was aware that when Hamilton's principle is applied, leading to $E = 0$, the term inside the square bracket on the RHS of (3) corresponds to the conserved quantity, or first integral, associated with the transformations (2).\footnote{In the case of field theory, the counterpart of the RHS is the total divergence of a current, so when Hamilton's principle is applied, (3) becomes a continuity equation (which Noether calls a ``law of conservation''), and a constant of the motion (Noether charge) only ensues when suitable boundary conditions are satisfied; see \cite{BH}, section 4. It is noteworthy that (i) this widely-known last step involving Gauss' divergence theorem is not mentioned by Noether and (ii) she also considers the case where the Euler-Lagrange expressions do not vanish, but she nontheless refers to laws of conservation in such cases; see footnote 15 in section 3 of \cite{Noether}} It is worth noting that Noether's theorem does not determine how many independent conservation laws follow from the invariance of the action under a $\rho$-parameter group of transformations. In relation to the full 10-parameter Galilean group, there are ten conservation laws, but for the single free particle only six are independent.\footnote{See \cite{Havas}, p. 151.}

Note also the role of $Q - \dot{q}T = [q'(t) - q(t)]/\epsilon$ on both sides of the Noether identity. Multiplied by $\epsilon$, it is what Noether denoted by $\overline{\delta q}$, and it is sometimes called the \emph{characteristic} of the symmetry (2),  We shall see below that defining new transformations (sometimes called \emph{contemporaneous}) with $Q$ replaced by $\tilde{Q} = Q - \dot{q}T$, and $T$ by $\tilde{T} = 0$, represents generally a quasi-symmetry that, assuming Hamilton's principle, leads to the same conserved quantity and obviously preserves the characteristic.\footnote{When $Q -\dot{q}T = 0$, the transformations will not be a strict symmetry; they will always be a quasi-symmetry associated with the trivial null conserved quantity. This will be made clear in the subsection 4.3,: the evolutionary form of these transformations is such that $Q' = T' = 0$.}

Now Noether's converse theorem, when applied to this simple case, starts from the identity (3) -- not the conservation law associated with the vanishing of the RHS of (3) -- and goes through the proof of the direct theorem ``in reverse order''. Noether presupposes a functional $\mathcal{L}(q(t), \dot{q}(t), t)$ along with its Euler-Lagrange expression (1), but does not presuppose a physical meaning for $Q$ and $T$. A series of reverse steps from (3) exploit the identity 
\begin{equation}
E\delta q = \delta \mathcal{L} - \frac{\mbox{d}}{\mbox{d}t}\left[\frac{\partial \mathcal{L}}{\partial \dot{q}}\delta q\right], 
\end{equation}
which follows from the calculus of variations where the first variation $\delta L$ arises from the variation $\delta t \equiv t' - t$ in the independent variable and $\delta q \equiv q'(t') - q(t)$ in the dependent variable. After integration, it is confirmed that $\int L\mbox{d}t$ is strictly invariant under the infinitesimal transformations (2), the factors $T$ and $Q$ thereby gaining their meaning, and $\delta q$ and $\delta t$ are shown to be linear in the parameter $\epsilon$. Thus Noether showed that the existence of a strict variational symmetry associated with $L$ holds if and only if (3) holds, or rather a generalisation thereof.\footnote{Noether remarked, however, that in the converse theorem, it is ``left moot'' whether in the case of transformations that depend on derivatives of the dependent variables (see below) the corresponding finite transformations form a group; see \cite{Noether}, section 3.} Note that \emph{the assumption that the Euler-Lagrange equations, and hence conservation laws, hold is nowhere used in the proof}.\footnote{This point is made particularly clearly in \cite{Bessel} and \cite{Havas}, pp. 158 and 161. In her \cite{K-S2}, Kosmann-Schwarzbach correctly states that in her converse theorem Noether started from ``linearly independent divergence relations'' (the higher dimensional generalisation of (3) above). She calls the result of applying the Euler-Lagrange equations to these relations ``conservation laws''. So it is somewhat misleading when she concludes that Noether had shown that
\begin{quote}
\ldots the existence of $\rho$ linearly independent conservation principles yields the infinitesimal invariance of [the Lagrangian] under a Lie algebra of infinitesimal symmetries of dimension $\rho$ \ldots 
\end{quote}}
In order to see if a strict variational symmetry can be correlated with a given conserved quantity arising from the Euler-Lagrange equations for the Lagrangian $\mathcal{L}$ of the simple kind we are considering, $T$ and $Q$ must be found such that (3) is satisfied and the conservation law corresponds to the vanishing of the RHS of (3). As Boyer noted in his careful 1967 study of Noether's theorem in the general case of field theory, the converse theorem ``gives no hint on the crucial question as to when a conserved current can be written in the form of the identity [corresponding to (3)]''.\footnote{See \cite{Boyer}, p. 457.}

However, in a little-known paper \cite{Dass} of 1966, Dass had proved a strengthened version of Noether's converse theorem. He showed that for a certain class of action integrals in field theory and for any divergenceless vector, there exists another such vector that differs from it by terms which vanish when the Euler-Lagrange equations hold, and for which the (generalised) version of (3) holds, leading to the existence of (quasi)symmetry transformations of the action. The restriction on the action is that the Lagrangian does not depend on the independent variables explicitly. As Havas pointed out in 1973\footnote{See \cite{Havas}. pp. 158, 162.}, this means that for our 1-dimensional mechanical case of $\mathcal{L} = \mathcal{L}\left(q(t), \dot{q}(t), t\right)$, any conservation law is associated with a (quasi)symmetry of the action as long as $\mathcal{L}$ does not depend explicitly on $t$\footnote{The case of a conservation principle for a particle undergoing frictional force in the second paragraph of footnote 13 above is not a counterexample because the Lagrangian depends explicitly on $t$.}, although it is not guaranteed in Dass' proof that the symmetry transformations form a group.
 
 Dass' theorem partially anticipates an even stronger general result in the more recent literature that allows for a version of the converse theorem that starts with a conservation law rather than a generalisation of the identity (3). In our special case of $\mathcal{L} = \mathcal{L}\left(q(t), \dot{q}(t), t\right)$, the quantity $P$ is conserved, i.e. $\mbox{d}P/\mbox{d}t = 0$ holds for all solutions of the  Euler-Lagrange equations $E = 0$, if and only if there is an `equivalent' conserved quantity $P'$ and a function $R = R(q, \dot{q})$ such that $ER = \mbox{d}P'/\mbox{d}t$. Olver calls this the \emph{characteristic form}, and $R$ the \emph{characteristic}, of the conservation law.\footnote{For Olver's proof of this result for more general Lagrangians, and for conservation laws of the form Div \emph{P} = 0, see \cite{Olver}, p. 270. It seems that the first reference to the characteristic form is due to Steudel in 1962; see \cite{Steudel62}. I am grateful to Peter Olver for bringing this paper to my attention.} Now note the similarity between this equation and (3) above, in which the analogue of $R$ is the characteristic $Q - \dot{q}T$ of the symmetry (2). Indeed, the modern version of Noether's combined forward and converse theorem rests on the key result  that a group of transformations defines a variational symmetry group related to given Lagrangian if and only if its characteristic is the characteristic of a conservation law for the associated Euler-Lagrange equations. A fuller account of the theorem (and the meaning of 'equivalence' above), will be given in section 4.3.
 
In the meantime, note that in 1970 Candotti, Palmieri and Vitale demonstrated a solution of a generalised version of (3) for strict symmetries in the case of a Lagrangian of the form $\mathcal{L}(q_i, \dot{q}_i, t)$, $i = 1, \ldots, n$ and conserved quantity $D$. In our case of a single degree of freedom, the solution is 
\begin{equation}
Q = - \frac{\partial D}{\partial \dot{q}}H^{-1} + \dot{q}T; \; \; \; T= \frac{1}{\mathcal{L}}\left[\frac{\partial \mathcal{L}}{\partial \dot{q}}\frac{\partial D}{\partial \dot{q}}H^{-1} - D\right],\end{equation}
where $H = \frac{\partial^2\mathcal{L}}{\partial\dot{q}\partial\dot{q}}$ is the Hessian of $\mathcal{L}$.\footnote{See \cite{CPV}, section IF. Note that their Euler-Lagrange expression and conserved quantity are the negative of ours. In the same paper, a related result for field theories is derived, for Lagrangian densities that again depend on the spatiotemporal coordinates, the fields and their first derivatives. The authors point out that these conditions on the Lagrangians and Lagrangian densities for particles and fields, respectively, can be relaxed ``somewhat'', but the derivations become more involved and less conceptually clear. Note finally that (5) will have singularities at the isolated points where the Lagrangian vanishes, as will equations (6) and (13) below.}

In anticipation of the discussion in section 4.3 concerning the generalised Noether theorem and its converse, it can be shown that the analogue of (5) in the case of quasi-symmetries is:
\begin{equation}
Q = - \frac{\partial D}{\partial \dot{q}}H^{-1} + \dot{q}T; \; \; \; T= \frac{1}{\mathcal{L}}\left[\frac{\partial \mathcal{L}}{\partial \dot{q}}\frac{\partial D}{\partial \dot{q}}H^{-1} - D -\mathcal{C}\right].
\end{equation}
These transformations correspond to the case of $\delta \mathcal{L} = \frac{\mbox{d}\mathcal{C}}{\mbox{d}t}\epsilon$.

\subsection{The free particle}

The simplest example of a Lagrangian of the above type is that associated with  the non-relativistic free particle in one dimension, $\mathcal{L} = \frac{1}{2} \dot{q}^2$.\footnote{Mass plays no role in Newton's first law of motion and is ignored here.} It can be used to exemplify the points we want to make, by looking at two elements of the Galilean symmetry group, namely boosts and time translations. Related remarks concerning scale symmetry are found in the Appendix.

While not strictly invariant under the one-parameter infinitesimal boost transformations
\begin{equation}
q' = q + \epsilon t;  \; \; \;  t' = t,
\end{equation}
(where the infinitesimal $\epsilon$ has dimension [$qt^{-1}$]) the free particle Lagrangian is quasi-invariant. Following the work \cite{Bessel} of Bessel-Hagen in 1921, it has been widely (though belatedly, as we shall see) recognised that Noether's direct theorem can be generalised to include cases of `quasi-' or `divergence' invariance, in the light of the `gauge' (divergence) freedom associated with Lagrangians.\footnote{Bessel-Hagen owed the idea of such generalisation to Noether herself; see \cite{Bessel} and \cite{K-S2}.} In the present example, the conserved quantity is $q - \dot{q}t$.\footnote{It is common to call the conserved quantities arising in Noether's theorem as first integrals (as Noether did herself), but sometimes a distinction is made between first integrals and constants of the motion (such as $q - \dot{q}t$) which depend on $t$.} It follows from the work \cite{CPV} of Candotti \emph{et al.} that there is a \emph{strict} variational symmetry (5) associated with this conservation law, in relation to the action associated with $\mathcal{L} = \frac{1}{2} \dot{q}^2$.\footnote{In this case, the strict symmetry, if it exists, is unique, but in general in higher dimensions uniqueness holds only up to the addition of a null divergence. See \cite{Olver} exercise 5.23(b), p. 372.} Curiously, Noether herself had already provided the explicit symmetry in her 1918 paper, without any reference to the issue of Galilean boost invariance in particle physics, but as an example of an invariance group $\mathcal{G}$ in which the  transformations of the dependent and independent variables may involve derivatives of the dependent variable (called \emph{generalised} symmetries by Olver\footnote{\cite{Olver}, p. 292.}); the fact that her theorem also applies to such symmetries was long overlooked.\footnote{In 1986 Olver (\cite{Olver}, p. 366) noted that generalized symmetries had been rediscovered many times; see also his \cite{Olver18} for their importance in soliton theory.}

In our notation, Noether's strict infinitesimal symmetry transformations are:
\begin{equation}
q' = q + (t - \frac{2q}{\dot{q}})\epsilon; \; \; \;
t' = t - \frac{2q}{\dot{q}^2}\epsilon,
\end{equation}
where the infinitesimal parameter $\epsilon$ again has dimension [$qt^{-1}$].\footnote{\cite{Noether}, section 3. I have not come across Noether's solution in the physics literature. In \cite{Sant} Santilli suggests (p. 342) the existence of non-standard symmetries associated with the conservation of centre of mass motion, but leaves the details to the reader.} This result surely \emph{casts doubt on the common claim that conservation of centre of mass motion is intrinsically linked to boost symmetry}.

We might further ask: is there a Lagrangian (expected to be a gauge transformed version of $\frac{1}{2}\dot{q}^2$) for which the boost transformations (7) are a strict variational symmetry? Again, Noether gave the answer: 
\begin{equation}
\mathcal{L} = \frac{1}{2}\left[ \dot{q}^2 - \frac{\mbox{d}}{\mbox{d}t}\left(\frac{q^2}{t}\right)\right],
\end{equation}
 and this can be verified using (3) with $Q = t$ and $T = 0$, which follow from transformations (7). Once more, the conserved quantity can be shown to be $q - \dot{q}t$.\footnote{In 1966, Denman \cite{Denman} showed that any Lagrangian for a classical particle moving in one direction which is strictly invariant under the Galilean transformations (7) must have the form $\mathcal{L}(w, t)$ where $w = q - \dot{q}t$. Noether's Lagrangian is $(q - \dot{q}t)^2/(2t^2)$.} This is all consistent with (5).\footnote{Noether's point in introducing the Lagrangian (9) was to show that in going from it with its strict symmetry (7), to the gauge-related Lagrangian $\mathcal{L} = \frac{\dot{q}^2}{2}$ with its strict symmetry (8), the term $\overline{\delta q} \equiv Q - \dot{q}T$ is preserved, and  the new transformations depend on derivatives of $q$ -- and that this is generally the case with such gauge-changes to the Lagrangian. We shall see the wider importance of this property of $\overline{\delta q}$ in the next subsection. For an interesting suggestion as to how Noether discovered generalised symmetries, see \cite{Olver18}.}
 
 Now the infinitesimal transformations corresponding to time translation
 \begin{equation}
q' = q ; \; \; \;  t' = t + \epsilon,
\end{equation} 
constitute both a quasi-symmetry of the Noether Lagrangian (9) as well as (famously) a strict variational symmetry of the standard Lagrangian $\mathcal{L} = \frac{1}{2} \dot{q}^2$, the associated conserved quantity in both cases being the Hamiltonian, in this case $\frac{1}{2} \dot{q}^2$.\footnote{The analogous case of spatial homogeneity hardly needs a Noetherian analysis; the fact that for the standard Lagrangian, $q$ is a cyclic, or ignorable variable, means that the Euler-Lagrange equation takes the form $\frac{\mbox{d}}{\mbox{d}t}\frac{\partial \mathcal{L}}{\partial \dot{q}} = 0$, which implies $\dot{q}$ is a constant of the motion.}  And notice that amongst the quasi-symmetries of the standard Lagrangian are 
\begin{equation}
q' = q - \frac{2q}{t}\epsilon; \; \; t'= t + \left( 1 - \frac{2q}{\dot{q}t}\right)\epsilon
\end{equation}
and 
\begin{equation}
q' = q - \dot{q}\epsilon; \; \; \; t' = t, 
\end{equation}
 the associated conserved quantity in both cases being again $\frac{1}{2} \dot{q}^2$. And from the result (5) of Candotti \emph{et al.}, we can calculate a strict symmetry associated with Noether's Lagrangian (9) and the same conserved quantity. The transformations are:
 \begin{equation}
q' = q - \dot{q}(1 -T)\epsilon; \; \; \; t' = t + T\epsilon; \; \; \; T = \frac{1}{\mathcal{L}}\left(\frac{q\dot{q}}{t} - \frac{1}{2}\dot{q}^2\right),
\end{equation}
which can be checked against (3), for $\mathcal{L}$ given by (9). (In all cases (10) - (13), $\epsilon$ has the dimension [$t$].)
 The existence of the quasi-symmetries (11) - (13) again \emph{flies in the face of the common assumption that conservation of energy is intrinsically related to the homogeneity of time}.\footnote{If an objection is made to the effect that strict symmetries are more fundamental than quasi-symmetries, then what is one to say about the boost symmetry (7) in relation to (8)?} 
 
  The suspicion that members of the pair (7) and (8) are in some sense equivalent, and similarly those of the quadruple (10), (11), (12) and (13), is borne out in the next subsection.

\subsection{The modern Noether map}

Being confined to strict symmetries, Noether's 1918 analysis arguably obscured the true nature of both the direct and converse theorems, as they are understood today. The key insight is that a bijective Noether map exists not between variational (quasi)symmetries and conservation laws, \emph{but between suitably identified equivalence classes of both}. To the best of my knowledge, the first proof of the existence of such a map, allowing for generalised symmetries (see above), is due to Luis Mart\'{i}nez Alonso in a regrettably little-known 1979 paper.\footnote{See \cite{LMA}; I am grateful to Peter Olver for alerting me to this paper, and for pointing out that it allows for generalised symmetries (private communication).} Then in 1986, Peter Olver independently proved the existence of the Noether map, again allowing for generalised symmetries.\footnote{See \cite{Olver86} and \cite{Olver}, p. 293.} In discussing post-Bessel-Hagen refinements to the direct theorem in his monumental 1986 monograph \emph{Applications of Lie Groups to Differential Equations}, Olver wrote:
\begin{quote}
\ldots Noether's theorem now provided a complete one-to-one correspondence between one-parameter groups of generalised variational symmetries of some functional and the conservation laws of its associated Euler-Lagrange equations. \ldots Recent results have further crystallised the roles of trivial symmetries and conservation laws in the Noether correspondence for totally nondegenerate systems, with the consequence that each nontrivial variational symmetry group gives rise to a nontrivial conservation law, and conversely.\footnote{\cite{Olver}, p. 293.}
\end{quote}

Olver expressed what I shall call the \emph{Mart\'{i}nez Alonso Olver (MAO) theorem} as follows:
\begin{quote}
\ldots if $\mathcal{L}$ is a nondegenerate variational problem, there is a one-to-one correspondence between equivalence classes of nontrivial conservation laws of the Euler-Lagrange equations and equivalence classes of variational symmetries of the functional [i.e. the action $\int \mathcal{L}\mbox{d}x$].\footnote{This is part of Theorem 5.42 in \emph{op. cit.} p. 328. Olver cites (p. 367) related results due to Vinogradov in 1984 \cite{Vino}, and to Steudel in 1966 \cite{Steudel}. In her \cite{K-S2} Kosmann-Schwarzbach refers to the `Noether-Olver-Vinogradov theorem', but I am unable to determine whether Vinogradov's theorem is equivalent to MAO.

It may be worth clarifying that the MAO theorem, being Lagrangian specific, does not imply that knowledge of some standard conservation laws determines a corresponding symmetry group of the dynamical equations of motion. After all, ``Since the conservation laws for a single particle are all a consequence of the constancy of its velocity, they are equally compatible with a law of motion (or indeed an action integral) invariant under the Lorentz group as with one invariant (up to a [surface term] only for the action integral) under the Galilei group.'' \cite{Havas}, pp. 157-8.}
\end{quote}
 
Throughout his book, Olver uses the group theoretical representation of a symmetry of a system of differential equations in terms of a vector field $\mathbf{v}$ over some open subset of the space of independent and dependent variables, where $\mathbf{v}$ incorporates infinitesimal generators of the relevant symmetry; the notion of a variational symmetry relative to some action  $\int \mathcal{L}\mbox{d}x$ is likewise defined in these terms. In what follows I will continue to use the more clunky but perhaps more familiar language of explicit transformations of the dependent and independent variables, and for purposes of illustration I will again use the simple example above of the free particle. The word `symmetry', when variational in relation to some action, will be taken to mean either a strict or a quasi-symmetry, unless otherwise specified.

In informal terms, Olver's condition of total nondegeneracy for a system of analytic Euler-Lagrange equations ensures local solvability of the system, so that it is neither over- nor under-determined. (The condition rules out systems with ``local" symmetries, the subject of Noether's second theorem.\footnote{The corresponding assumption in Martinez Alonso's paper \cite{LMA} is his `normality' condition on a system of partial differential equations. Note that by a symmetry being ``local", I mean one in which the transformations of the independent and dependent variables are spacetime dependent, unlike in (2) above, when the parameter $\epsilon$ is ``global'', i.e. does not depend on $t$. In the mathematics literature, including Olver's 1986 monograph, the term ``local'' applies to any symmetry if its infinitesimal generator depends only on the independent and dependent variables and their derivatives, i.e. the jet coordinates.  According to this definition, both Noether's first and second theorems concern local symmetries.

In order to avoid possible confusion, theories with local (in our sense) symmetries, such as gauge-invariant Maxwellian electrodynamics formulated in terms of 4-potentials, are nonetheless subject to the MAO theorem as long as they also possess symmetries which are global. The first application of Noether's first theorem to electrodynamics was given in 1921 by Bessel-Hagen \cite{Bessel}.}) Two conservation laws are equivalent according to Olver if they differ by a trivial conservation law, which is a linear combination of trivial laws of the first and second kind. An example of equivalence of the first kind for the free Newtonian particle is that between $\frac{\mbox{d}}{\mbox{d}t}\left(\frac{1}{2}\dot{q}^2\right) = 0$ and $\frac{\mbox{d}}{\mbox{d}t}\left(\frac{1}{2}\dot{q}^2 + \ddot{q}\right) = 0$, since $\ddot{q}$ vanishes on-shell. An example of a trivial law of the second kind for higher dimensional systems is Div $P$ = 0 when $P$ is a total curl, i.e. the law is a mathematical identity, and holds off-shell.\footnote{\cite{Olver}, pp. 268 -270.} Finally, two symmetries are equivalent if they differ by a trivial symmetry. In order to understand what Olver means by a trivial symmetry, a short but revealing detour is needed.

A notable feature of any symmetry in relation to a given system of differential equations -- whether or not there is a Lagrangian for which they are Euler-Lagrange equations -- is that if the associated transformations are not restricted to just the dependent variables, then there is a procedure for generating from them a symmetry that does have this property, associated moreover with the same first integral.\footnote{For the general case, see, e.g. \cite{Boyer}, p. 451.} Recall that if the transformations (2) are a quasi-symmetry in relation to the action  $\int\mathcal{L}(q, \dot{q}, t)\mbox{d}t$ then
\begin{equation}
E(Q -\dot{q}T) = \frac{\mbox{d}}{\mbox{d}t}\left[-\frac{\partial \mathcal{L}}{\partial \dot{q}}\left(Q -\dot{q}T\right)  - \mathcal{L}T + \mathcal{C}\right], 
\end{equation}
where $\delta L = \frac{\mbox{d}}{\mbox{d}t}\mathcal{C}\epsilon$. Now consider the transformations defined by 
\begin{equation}
q' = q + \tilde{Q}\epsilon; \; \; \; \tilde{Q} = Q - \dot{q}T; \; \; \;  t' = t.
\end{equation}
Olver calls these contemporaneous transformations the \emph{evolutionary representative, or form}, of (2), preserving the characteristic of (2). 
If Noether's 1918 converse theorem also holds for quasi-symmetries, then it is easy to confirm that the transformations (15) are a quasi-symmetry of the same action, with the same conservation law, but with $\mathcal{C}' = \mathcal{C} - \mathcal{L}T$.  Indeed, it is a theorem that a dynamical symmetry of a given system of differential equations is a variational symmetry in relation to the action $\int \mathcal{L}\mbox{d}x$ if and only if its evolutionary representative is too.\footnote{see \cite{Olver}, Proposition 5.36, p. 325.} Note that it follows immediately from both equations (5) and (6) that whether the symmetry (2) is a strict or a quasi-symmetry, $\tilde{Q}$ takes the form:
\begin{equation}
\tilde{Q} = -\frac{\partial D}{\partial \dot{q}}H^{-1},
\end{equation}
 so apart from the conserved quantity $D$, $\tilde{Q}$ depends only on the Hessian of the Lagrangian.

Now Olver calls a symmetry of a system of differential equations \emph{trivial} if \emph{its evolutionary form has coefficients which vanish on-shell}. In our case, this means that the symmetry (2) is trivial when the coefficient of $\epsilon$ in (15) vanishes on-shell, such as in the case of $q' = q + \ddot{q}\epsilon', \; t' = t$. Olver writes:
\begin{quote}
Two generalized symmetries $\mathbf{v}$ and $\tilde{\mathbf{v}}$ are called \emph{equivalent} if their difference $\mathbf{v} - \tilde{\mathbf{v}}$ is a trivial symmetry of the system. This induces an equivalence relation on the space of generalized symmetries of the given system; moreover, we will classify symmetries up to equivalence so by a \emph{symmetry} of the system we really mean a whole equivalence class of generalized symmetries, each differing from the other by a trivial symmetry.\footnote{\cite{Olver}, p. 298. Note that a dynamical symmetry may be equivalent to a variational quasi-symmetry but not itself be one. Recall the quasi-symmetry (7); an equivalent dynamical symmetry obtained by adding $\ddot{q}$ to the $q$ transformation is no longer a variational symmetry. See \cite{Olver} exercise 5.22, p. 372.} 
\end{quote}

Olver gives some examples related to the heat equation (for flow in a one-dimensional rod with unit diffusivity):
\begin{equation}
\frac{\partial u}{\partial t} - \frac{\partial^2u}{\partial x^2} = 0,
\end{equation}
in terms of the following infinitesimal symmetries:
\begin{equation}
x' = x, \; \; t' = t + \epsilon, \; \;u'=u,
\end{equation}
\begin{equation}
x' = x, \; \; t' = t, \; \; u' = u -\dot{u}\epsilon',
\end{equation}
and 
\begin{equation}
x' = x, \; \; t' = t, \; \; u' = u - \frac{\partial^2u}{\partial x^2}\epsilon'',
\end{equation}
where the infinitesimals $\epsilon$ and $\epsilon'$ have dimension [$t$], and $\epsilon''$ has dimension [$x^2$].\footnote{Emphasis on the heat equation in this context is perhaps unfortunate, because as a scalar evolution equation it does not have a Lagrangian formulation. But there is a twist in the tale; see Appendix A below.}  As Olver states, (19) is the evolutionary form of time translation (18), and he claims that all three symmetries are equivalent  -- and ``for all practical purposes determine the self-same symmetry group''.\footnote{\emph{Ibid}. The notion of equivalence here needs nuancing. The time translation (18) has, for an isolated subsystem of the universe, a familiar and straightforward operational rendering that (19) and (20) do not, related to the homogeneity of time. Olver's notion of equivalence is formal, and meaningful only in relation to a given set of dynamical equations. Similar remarks hold for the ``equivalences" of certain transformations discussed in section 4.2 for the free particle; see below.} Indeed, subtracting (20) from (18), as well as (20) from (19), result in symmetries whose evolutionary forms are trivial on-shell.
But subtracting (19) from (18) results in a symmetry whose evolutionary form is trivial \emph{because its coefficients vanish identically}, and thus vanish off-shell. (A treatment of these results in Olver's vectorial notation is given in Appendix A.) In fact, it is easy to prove that the difference between any symmetry and its evolutionary form has this property. 

Now it is noteworthy that before Olver gives the above definition of triviality, when introducing the notion of the evolutionary representative of a symmetry he claims that the two are ``essentially the same symmetry''.\footnote{\cite{Olver}, p. 297.} Thus there are, as in the case of conservation laws, \emph{two types of triviality} in the case of evolutionary symmetries that Olver is effectively assuming: on-shell and off-shell. Recognition of this point does not, however, change the details of the proof of the MAO theorem.\footnote{As mentioned in section 4.1 above, Olver's proof rests on the key result that a group of transformations defines a variational symmetry group related to given Lagrangian if and only if its characteristic is the characteristic of a conservation law Div $P = 0$ for the associated Euler-Lagrange equations. In defining an equivalence class of variational symmetries the only relation between characteristics that is of interest is the on-shell relation: symmetries related by off-shell equivalence automatically have the same characteristic. (A similar situation holds for the proof of Martinez Alonso.) I am grateful to Peter Olver for clarifying this matter (private communication). Note that by admitting off-shell equivalence, an equivalence class of symmetries gains members but the cardinality of the set of equivalence classes is not altered, and remains the same as that of the equivalence classes of conservation principles.} 

We return now to the Lagrangian formulation of the dynamics of the free particle. It is easy to confirm that the boost transformations (7) are the evolutionary representative of Noether's transformations (8), and that the transformations (12) are the evolutionary representative of (10), (11) and (13). So (7) and (8) are equivalent in the off-shell sense and the same goes for  (10), (11) (12) and (13) -- and note that the formula for $\tilde{Q}$ in (16) is consistent with (7) and (12) given the associated first integrals $q -\dot{q}t$ and $\frac{\dot{q}^2}{2}$, respectively. 

Note too that whether two symmetries are equivalent because one is the evolutionary form of the other is insensitive to the gauge chosen for the relevant Lagrangian.
Now recall that the MOA theorem above defines equivalence classes of variational symmetries \emph{relative to a given action}. So far we have seen that transformations (7),(8), and (10)-(13) are variational symmetries of the action associated with the standard Lagrangian $\mathcal{L} = \frac{1}{2} \dot{q}^2$, and the transformations (7) and (10) are variational symmetries of the action associated with gauge-related Noether Lagrangian (9). Consistency with the MOA theorem and the off-shell definition of equivalence demands that all of these transformations should be variational symmetries of both Lagrangians, with (7) and (8) having associated conserved quantity  $q -\dot{q}t$, and (10) - (13), having $\frac{\dot{q}^2}{2}$. 

Calculations show that this is so, and summary details are provided in Appendix B.  It is seen that neither of these Lagrangians is strictly invariant under both boost transformations and time translations. In a recent paper \cite{Olver20}, Olver has demonstrated that there is no non-constant first-order Lagrangian for the non-relativistic free particle that is strictly invariant with respect to the full Galilean group. In the same paper he also provided a necessary and sufficient condition, using cohomology theory, for every divergence-invariant Lagrangian related to a connected Lie group of point transformations to be locally equivalent to a strictly invariant Lagrangian.

\section{Global space-time symmetries}

If the original question of explanatory priority cannot be settled in a straightforward way in the light of the modern bijective Noether map, there may be other reasons why symmetries are routinely given privileged status. 

The quotations from Landau and Lifshitz, Zee, and Ryder in section 2 above refer specifically
to space and time symmetries. Properties of space such as homogeneity and isotropy, of time such as homogeneity, and properties of space-time such as boost invariance, seem to transcend the physics of any one type of non-gravitational interaction. It is such universality that might suggest that these properties are primitive attributes of the very space-time manifold on which different field theories are written, and more fundamental than conservation principles.\footnote{The pertinent issue here is universality, not intuitive appeal. The boost claim is surely less intuitive than the homogeneity claims. As regards the latter, Lewis Ryder (\cite{Ryder}, section 3.2) wrote: ``If it were not for these conditions, it is obvious that science itself would be impossible.'' More complicated maybe, but arguably not impossible; physics has accommodated the violation of some space-time symmetries, as we see below.}
But note that as late as 1900, H. A. Lorentz would extol two principles as putative universal constraints on theories: the second law of thermodynamics and the conservation of energy.\footnote{See \cite{Frisch}, section 3. In } It was Einstein's special theory of relativity that put the spotlight on the role of symmetries \emph{qua} constraints, and we return to this development in the following section.

At this point it is worth reminding ourselves that some natural-looking properties of space and time have turned out \emph{not} to be universally valid. The weak interactions ``have little respect for symmetries''\footnote{\cite{BM-V}, section 3.}, at least of the discrete kind. The experimental violation of parity ($P$, or space-inversion) in 1957 came as a bolt out of the blue; the experimental violation of time inversion symmetry announced in 2012 was less surprising, given the violation of $CP$ symmetry (a combination of $P$ and charge conjugation $C$) in the weak interactions in 1964 and 2001.\footnote{For details see \cite{Lees} and \cite{BM-V}.} These results are hard to reconcile with the claim that space-time symmetries have a kind of metaphysical necessity: cherry picking is not allowed. It makes sense then that searches still take place for possible violations of Lorentz covariance.\footnote{See \cite{Will}, section 2.1.2.} 

It is also helpful to look at the significance of these global spacetime symmetries from the perspective of Einstein's general theory of relativity (GTR). None of these symmetries emerges from inspection of Einstein's field equations, whose symmetry group is the local diffeomorphism group. They arise from the specific way matter fields couple to the metric field as determined by the strong (Einstein) equivalence principle, which brings the Poincar\'{e} group of relativistic boosts, rotations and translations, and hence the local validity of special relativity, into the story. Space, for instance, is not intrinsically homogeneous nor isotropic from the perspective of the gravitational (metric) degrees of freedom; these properties emerge from the Euclidean subgroup of the Poincar\'{e} group associated with the matter field equations when expressed in the appropriate local freely falling frames. The universality of local Poincar\'{e} covariance for the diverse non-gravitational interactions is one of the remarkable features of the strong (or Einstein) equivalence principle in GTR. Such covariance is a property of the relevant matter field equations, and in this sense the same conclusions can even be applied to the nature of space-time in special relativity. Finally, even such continuous, global space-time symmetries are, from the point of view of GTR, only approximately valid. They hold in circumstances in which the effects of space-time curvature are negligible.\footnote{This brief discussion of the strong equivalence principle in GTR is over-simplified; for more extensive discussion see \cite{RBL}.}

\section{A brief ode to symmetries}

\subsection{Pragmatic considerations}

(i) It is striking that, despite its fame in physics, Noether's theorem was largely overlooked by physicists for decades. Olver wrote in 1986:
\begin{quote}
 \ldots by 1922 [the year following Bessel-Hagen's version of the direct theorem for quasi-symmetries] all the machinery for a detailed, systematic investigation into the symmetry properties and consequent conservation laws of the important equations of mathematical physics was available. Strangely enough, this did not occur until quite recently.\footnote{\cite{Olver} p. 288. Yvette Kosmann-Schwarzbach has provided an extensive treatment of the history of the reception of Noether's 1918 paper in \cite{K-S1}, a pr\'{e}cis of which is found in her \cite{K-S2}. For related remarks see also \cite{Olver18}.}
 \end{quote}
 As Olver noted, an important event was the publication in 1951 by E. L. Hill of a review of (a special case of) Noether's direct theorem in \emph{Reviews of Modern Physics}. In this paper, Hill lamented that 
\begin{quote}
 Despite the fundamental importance of this theory
there seems to be no readily available account of it
which is adapted to the needs of the student of mathematical physics, while the original papers [of Klein, Noether and Bessel-Hagen] are not readily accessible.
\end{quote}
As late as 1964, Tassie and Buchdahl published a paper \cite{TB} extending the direct theorem to quasi-symmetries, in apparent ignorance of the 1921 work of Bessel-Hagen. An English translation of Noether's 1918 paper was not published until 1971, at which time applications of her theorem in mathematical physics were still thin on the ground. A translation of Bessel-Hagen's 1921 paper, which applied the generalised Noether theorem to electrodynamics for the first time (including its conformal symmetry) only appeared in 2006.   
In Olver's comments on the history of the subject, he noted that  important applications of conservation principles and Noether-type identities in elasticity, scattering theory and optics occurred before the connection with Noether's theorem was realised.\footnote{See \cite{Olver}, p. 288.} 

It might help to pause for a moment on the case of elasticity. A major advance in the treatment of defects (dislocations, point defects, interfaces and crack tips) within elastic media emerged in the work of J. D. Eshelby between 1951 and 1956.\footnote{See \cite{Esh51} and \cite{Esh56}; Eshelby's contributions to the field extended well beyond these initial papers.} Eshelby was able to prove the existence of a stress-energy tensor, and a related conservation law, from which the force on the defect can be calculated, in analogy to the role of the Maxwell stress tensor in electrostatics in establishing the force on a charge.\footnote{A very clear treatment of the application of Eshelby's method in the cases of pressure in an interface and force on a static defect is found in \cite{Sutton}, Chapter 8.} The 1968 work of G\"{u}nther and the independent work by Knowles and Sternberg in 1972 had together established, using Noether's theorem, the relevant conservation law on the basis of spatial translation symmetry of an elastically homogeneous material, a second conservation law that holds in the case of homogeneity and isotropy\footnote{The homogeneity and isotropy of an elastic medium  are usually defined in terms of the corresponding properties of the elastic constants of the medium, but sometimes the defects themselves are regarded as inhomogeneities. From the point of view of Noether's theorem, what matters is whether the Lagrangian density itself (the negative of the elastic energy density) is symmetric in these ways.}, and a third law in relation to the assumption of scale invariance.\footnote{See \cite{Gunth} and \cite{Knowles}. It is worth noting two features of this application of Noether's theorem. First, in elastostatics, time obviously makes no appearance; the infinitesimal transformations are defined for the spatial coordinates and the displacement vector field, and the conservation laws does not yield  constants of the motion in the usual temporal sense of mechanics. Indeed they represent a counterexample to the claim that ``Noether's theorem concerns the way particles behave \emph{under temporal evolution}'' (\cite{Sch}, p. 52). (Recall point 5 in section 3 above.) Second, the theorem holds only in homogeneous regions of space inside the solid body which have no defects, though it allows for forces on defects to be calculated. For subsequent developments involving such conservation laws, see \cite{Rice}.} In 1984, Olver found further undetected symmetries of the equations of linear elasticity, associated with new conservation laws.\footnote{See \cite{Olver3} and \cite{Olver4}. A summary of some of these applications of Noether's theorem is found in \cite{Olver}, Example 4.32, pp. 281-283. A recent paper \cite{Fed} provides an explicit relation between Eshelby's inclusion theory  and Noether's theorem, and cites several more papers on the role of Noether's theorem in the theory of elasticity published since Olver's work.} 

Noether's theorem played two roles in this development of the theory of defects in elastic media. First, here is Eshelby himself in 1975:
\begin{quote}
The normal theory of elasticity recognizes nothing which corresponds with the force on a defect. \dots But in fact the appropriate concept has been to hand ever since the appearance of a paper by Noether \ldots in 1918, in the form of the energy-momentum tensor which the elastic field possesses in common with every field whose governing equations are derivable from a variational principle, and some for which they are not.\footnote{\cite{Esh75}, p. 322.}
\end{quote}
Eshelby had not needed Noether's theorem for the discovery of his stress-energy tensor, but by giving in 1975 a Lagrangian treatment of elasticity, he wanted to show that the theory was amenable ``to the standard results of the classical part of general field theory''.\footnote{In this way Eshelby hoped to attract the interest of applied mathematicians in the stress-energy tensor, the lack of which (the work of Gunther and Knowles and Sternberg being exceptions) he attributed tentatively to ``the artificial separation which has grown up between applied mathematics and theoretical physics''. \cite{Esh75}, p. 323.} 

But more importantly for our purposes, the use of Noether's theorem led to the discovery of new conservation laws for elastic media. And here a discrepancy between symmetries and conservation laws is apparent in practice: it seems that it is easier in general to make progress by using Noether's direct theorem than by using its converse. In the case of elasticity, the symmetries follow directly from the nature of the Lagrangian which in turn reflects the assumed properties of the elastic medium under consideration. 

(ii) Mention should be made of the fact that knowledge of first integrals plays a role in the possible solution of equations of motion by quadrature (integration), including numerical integration.\footnote{For mechanical systems, the original meaning of the term (first) integral was not a constant of the motion but an equation of the form $\frac{\mbox{d}}{\mbox{d}t}f(q_i, \dot{q}_i, \ddot{q}_i, ..., t) = 0$, which can be solved by integration.} For instance in the case of $\mathcal{L}(q, \dot{q}, t)$, knowledge of a conserved quantity allows for integration of the Euler-Lagrange equation completely by quadratures.  For the more general case involving $n$-th order variational problems, knowledge of a one-parameter group of variational symmetries allows for reduction of the associated Euler-Lagrange equations of order 2$n$ to those of order 2$n -2$.\footnote{See \cite{Olver}, Theorem 4.17, p. 262.} In 1986 Olver wrote that ``Noether's method is the only really systematic procedure for constructing conservation laws for complicated systems of partial differential equations'',\footnote{\cite{Olver} p. 246.} although more recently computer algorithms have been developed to generate conservation laws which are independent of variational considerations.\footnote{For details see \cite{Wolf}.} 

(iii) Finally, following the seminal work of Wigner in 1939 \cite{Wigner}, there is widespread acceptance by physicists of the claim that the very properties of elementary particles are grounded in continuous symmetry groups. Mass and spin, for example, are 
represented by Casimir invariants of the Poincar\'{e} group. In the words of Steven Weinberg, such properties ``are what they are because of the symmetries of the laws of nature.''\footnote{\cite{Wein}, p. 138n, also cited in \cite{Sch}, which provides a recent philosophical analysis of such claims.}

I do not wish to assert that any of the pragmatic considerations in this subsection are likely to be the motivation for the claims quoted in section 2 above. But these considerations seem to support the notion that Noether's direct theorem has proved more useful in mathematical physics than its converse, as well as to underline the prominent role of symmetries in the articulation of particle physics.

\subsection{Heuristics} 

The prominent heuristic role symmetries have played in twentieth century physics must surely be part of our account. This story is well-known, but to my knowledge has not been stressed in discussions of the explanatory arrow in Noether's theorem, at least in the philosophical literature. So the reader may forgive a brief recap.

Einstein's 1905 derivation of the Lorentz transformations rested on two fundamental symmetry principles: the relativity principle (dynamical equivalence of inertial frames) and the isotropy of space, alongside the  postulate governing the constancy of the speed of light with respect to the ``resting'' frame.\footnote{For the role specifically of spatial isotropy in Einstein's derivation, see \cite{PR}, section 5.4.3.}  The justification of all these principles did not rest, for Einstein, on any \emph{a priori} notions about the structure of space and time, but was  based on ``plenty of experiential knowledge'' related to mechanics and electrodynamics.\footnote{Einstein to Amiet, Dec. 17, 1947; The Albert Einstein Archives at the Hebrew University of Jerusalem, 25-335.} In 1940, he would stress that the theory of special relativity could be summarised in one principle: ``all natural laws must be so conditioned that they are covariant with respect to Lorentz transformations''.\footnote{\cite{Ein40}, p. 490.} This allowed Einstein to say that the theory transcended Maxwell's equations, and what he saw as the awkward emphasis on the role of light in his 1905 formulation.\footnote{For further discussion, see \cite{PR}, chapter 8, and especially \cite{Giov}, section 4.1.}  Special relativity is essentially a constraint in the sense that a symmetry is being imposed on  the fundamental equations of \emph{all} the non-gravitational interactions. This amounts to only the second time that a constraint, or set of constraints, on fundamental physics have been given the honorific title of a theory; the first was thermodynamics.\footnote{The difference between the 1940 formulation (as distinct from the 1905 one) and thermodynamics is that now the constraint is no longer purely phenomenological, and hence special relativity ceases to be a strict ``principle'' theory in Einstein's terminology. See \cite{PR}, section 8.4.1.}

One of the most remarkable methodological trends in modern physics has been the \emph{a priori} use of symmetry principles to constrain the action principles of the non-gravitational interactions in quantum electrodynamics (QED) and particle physics. It is what A. Zee called the new paradigm of ``symmetry $\rightarrow$ action $\rightarrow$ experiments'' in fundamental physics.\footnote{\cite{Zee2}, p. 457; see also \cite{Zee1}, chapter 6 and \cite{Olver18}.} Indeed the use of the Lagrangian formalism has become virtually \emph{de rigeur} in these fields (though less so in axiomatic, algebraic quantum field theory\footnote{See, e.g., \cite{Orz}, where it is argued that it is the converse of Noether's theorem that is valid.}) largely because it is so friendly to the imposition of symmetry constraints in comparison to conservation laws.\footnote{In Pauli's magnificent 1921 monograph \cite{Pauli21} on relativity theory, the English translation of which was published in 1958, he wrote (\cite{Pauli}, p. 201) that ``it is not at all self-evident, from a physical point of view, that the physical laws should be derivable from an action principle.'' Such qualms are rare today, at least in relation to fundamental physics.} Symmetries can be said to be worn on the sleeves of the Lagrangian (which is not to say that given a Lagrangian, \emph{all} of its symmetries ``leap out at you''\footnote{See \cite{Zee2}, p. 457, for this point.}). And the issue goes beyond the special relativistic constraint (Lorentz invariance).\footnote{According to Peter Havas, the heuristic role of \textit{universal} Lorentz invariance ultimately finds its justification in terms of belief in the standard conservation principles:
\begin{quote}
The fundamental importance of the invariances lies in their role in the development of physical theories, in the help they may provide in our search for ``all the laws of nature''. Thus, once physicists had come to trust the special theory of relativity, the requirement of invariance under the proper orthochronous Lorentz group served as a selection criterion in limiting the number of theories acceptable for further scrutiny as possible candidates for a correct description of parrticular classes of phenomena. If we then further require that all the laws of nature should be derivable from a \emph{single} Lorentz invariant \ldots variational principle, we are automatically provided with a \emph{universal} set of ten conservation laws which can be readily interpreted in terms of familiar physical concepts.

It is precisely the strength of the physicists' belief in the universal validity of such a set of conservation laws which motivates their preference for a single variational principle invariant under a universal group in the first place. \cite{Havas}, p. 160.
\end{quote}} An equally, if not more, important actor in the story is the Weyl-Yang-Mills \emph{gauge principle} involving internal symmetries.

QED is based on a Lagrangian
\begin{equation}
\mathcal{L}_{\mbox{QED}} = \mathcal{L}_{\mbox{Dirac}} + \mathcal{L}_{\mbox{Maxwell}} + \mathcal{L}_{\mbox{int}},
\end{equation}
coupling the Dirac field with the Maxwell field. This Lagrangian can be considered the result of taking the Dirac Lagrangian for the free electron, which is invariant under global phase transformations associated with the abelian U(1) group, and replacing it by one which is invariant under local (space-time dependent) phase or ``gauge'' transformations. Needless to say, the resulting appearance of a 4-vector potential (or connection in the language of fibre bundles), with its appropriate compensating transformations, does not mean anything physically interesting \emph{per se}. It is the further requirements that such a ``gauge'' field (a) be dynamical and (b) give rise to a locally invariant kinetic energy term in the Lagrangian that depends only on the the 4-potential and its derivatives, and (c) interact with the Dirac field by way of ``minimal coupling'', i.e, by replacing derivatives of the Dirac field by a suitable covariant derivative in the interaction term in the Lagrangian,\footnote{Note that the minimal coupling assumption is appropriate only if the free Dirac Lagrangian is chosen to be ``minimal'' with respect to Lagrangian gauge transformations; see \cite{Sak}, p. 183.} that brings empirical clout to the procedure such that the 4-potential gains its familiar electrodynamic currency.\footnote{It is also necessary in this procedure to assume renormalizability as well as the invariance of the Lagrangian under time reversal or parity; see \cite{Pesk}, section 15.1. It is noteworthy that in his excellent 2004 monograph \emph{The Geometry of Physics} Theodore Frankel gives the misleading impression that step (a) in the gauge principle is a requirement of quantum mechanics, rather than an independent postulate (\cite{Frank}, p. 535). A more accurate spin on the significance of the 4-vector field is given in \cite{Blund}, p. 128:
\begin{quote} 
It only exists in our description because we've invented it to satisfy our demand for a locally invariant theory, but if such ambitions have any grounding in reality, then the [4-vector] field should have dynamics of its own! 
\end{quote}
See also \cite{Obj}, section 2(a). An amusing article by Heras \cite{Heras} shows that applying the gauge principle to the Schr\"{o}dinger equation for a free electron, without imposing the analogue of (b) above, is consistent with an elliptical variant of Maxwell's equations with Euclidean, rather than Lorentzian, symmetry, and hence possessing no propagating solutions and no invariant speed.} In 1954, Yang and Mills \cite{YM} generalised this ``gauge principle" to a system with global symmetry (isotopic spin rotation for nucleons) associated with a nonabelian group. The rest of the story is encapsulated in the words of Zee:
\begin{quote}
In the late 1960s and early 1970s, the electromagnetic and weak interactions were unified into an electroweak interaction, described by a nonabelian gauge theory based on the group $SU(2)\otimes U(1)$. Somewhat later, in the early 1970s, it was realised that the strong interaction can be described by a nonabelian gauge theory based on the group $SU(3)$. Nature literally consists of a web of interacting Yang-Mills fields.\footnote{\cite{Zee2}, p. 361. The original problem facing Yang-Mills theory is that it gives rise to massless spin-1 particles which, apart from the photon, are not observed. In the eventual electroweak theory, the Yang-Mills particles acquire mass through the Higgs mechanism, and in quantum chromodynamics, the unobservability of the particles (gluons) is explained through the phenomenon of asymptotic freedom; see, e.g., chapters VII.2 and VII.3 (\emph{op. cit.}).}
\end{quote}

It may be unclear why the gauge principle works when it does, and for some commentators the fact that the physics of the non-gravitational interactions is redolent of gauge symmetry seems to represent an awkwardness in the standard model.\footnote{See, e.g., \cite{Zee2}, p. 456.} Be that as it may, the heuristic role of symmetry in the development of post-19th century physics, including string theory, has no comparable historical precedents.\footnote{The one, isolated pre-Einstein case is that of Christiaan Huygens, who in 1656 used the relativity principle as a postulate in deriving his non-Cartesian theory of collisions; an insightful account of this episode is given by Barbour in \cite{Barb}, section 9.4. Barbour mentions on p. 470 the intriguing possibility, first mooted by Martin Klein, that Einstein may have been influenced by the account of Huygens' theory of collisions in Mach's \emph{Mechanics}, a book which he ``read avidly''. (A fable concerning a 1705 anticipation of Einstein's 1905 logic, in which Albert Keinstein uses the relativity principle to derive the Galilean transformations, is found in \cite{PR}, chapter 3.)

 Justification for the claim that Einstein's route to his 1915 theory of gravity, general relativity, was motivated (\emph{inter alia}) by a symmetry principle, namely general covariance, is relatively problematic. Not only did Einstein erroneously think, until at least 1918, that general covariance was a generalisation of the relativity principle, he also abandoned the principle between 1913 and early 1915. For details, see \cite{PR} Appendix A, and especially \cite{Janssen}, section 3.}  It has also led to the use of advanced geometrical and group theoretical methods in particle and condensed matter physics. 

It is hard to believe that this development has not, to some extent, influenced the conventional reading of Noether's theorem, at least amongst physicists since the 1970s. While, again, it does not strictly justify the claim that symmetries have explanatory priority in Noether's theorem, recognition of the important heuristic role of symmetries in the standard model may be a contributing factor behind the relative unfamiliarity of the converse Noether theorem and its implications.

\bibliographystyle{plain}

\medskip

\begin{thebibliography}{9}

\bibitem{And} N. Anderson and A.M. Arthurs, A variational principle for Maxwell's equations, International Journal of Electronics Theoretical and Experimental, 45:3, (1978) 333-334, DOI: 10.1080/00207217808900916
\bibitem{Barb} J. Barbour, The Discovery of Dynamics, Oxford University Press, New York, 2001.
\bibitem{BM-V} J. Bernab\'{e}u and F. Mart\'{i}nez-Vidal, Colloquium: Time-reversal violation with quantum-entangled \textit{B} mesons, Reviews of Modern Physics, 87(1), (2015) 165-182;  DOI: 10.1103/Rev Mod Phys.87.165
\bibitem{Bessel} E. Bessel-Hagen, \"{U}ber die Erhaltungss\"{a}tze der Elektrodynamik. Mathematische Annalen 84 (1921), 258-276. English translation by M. Albinus and N.H. Ibragimov: On Conservation Laws of Electrodynamics, in Archives of ALGA, vol. 3 (2006) 33-51. ALGA Publications ISSN 1652-4934. \url{https://www.researchgate.net/publication/311456950_Archives_of_ALGA_vol_3}.
\bibitem{BT} J.W. Blaker, and M.A. Tavel, The Application of Noether's Theorem to Optical Systems, American Journal of Physics 42, 857-861. (1974); doi: 10.1119/1.1987878
\bibitem{Boyer} T. H. Boyer, Continuous Symmetries and Conserved Currents. Annals of Physics, 42, 445-466  (1967).
\bibitem{BB} K. Brading and H.R. Brown, Symmetries and Noether's Theorems, in Symmetries in Physics: Philosophical Reflections, K. Brading and E. Castellani (eds.), Cambridge University Press, 2003; pp. 89-109.
\bibitem{Obj} H.R. Brown,  Aspects of objectivity in quantum mechanics, in From Physics to Philosophy, J. Butterfield and C. Pagonis (eds.), Cambridge University Press, 1999; 45-70. \url{http://philsci-archive.pitt.edu/223/1/Objectivity.pdf}. 
\bibitem{PR} H.R. Brown, Physical Relativity. Space-time structure from a dynamical perspective, Clarendon Press, Oxford (2007).
\bibitem{BH} H. R. Brown and P. Holland, Dynamical vs. variational symmetries: Understanding Noether's first theorem, Molecular Physics 102, (11-12 Spec. Iss), 1133-1139 (2004). \url{http://philsci-archive.pitt.edu/2914/1/MolPhys04.pdf}.
\bibitem{BH2} H.R. Brown and P.R. Holland, Simple applications of Noether's first theorem in quantum mechanics and electromagnetism, 2004, American Journal of Physics,
72, 34-39. \url{http://arxiv.org/quant-ph/0302062}. 
\bibitem{Butt} J. Butterfield, On Symmetry and Conserved Quantities in Classical Mechanics, in Physical Theory and its Interpretation, W. Demopoulos and I. Pitowsky (eds.), Springer 2006, pp. 43-99.
\bibitem{CPV} E. Candotti, C. Palmieri, and B. Vitale, On the Inversion of Noether's Theorem in Classical Dynamical Systems,
American Journal of Physics 40, 424-429 (1972); doi: 10.1119/1.1986566.
\bibitem{Dass} T. Dass, Conservation Laws and Symmetries. II, Physical Review 150(4) (1966), 1251-1255.
\bibitem{Denman} H.H. Denman, Invariance and Conservation Laws in Classical Mechanics. II.,
Journal of Mathematical Physics, 7(11) (1966), 1910-1915.
\bibitem{Ein40} A. Einstein, Considerations concerning the Fundaments of Theoretical Physics, Science 91, (1940) 487-492.
\bibitem{Esh51} J.D. Eshelby, The force on an elastic singularity,  Philosophical Transactions of the Royal Society of London A 244, 87-111 (1951) \url{https://doi.org/10.1098/rsta.1951.0016}. 
\bibitem{Esh56} J.D. Eshelby, 
The Continuum Theory of Lattice Defects, Solid State Physics 3, 79-144 (1956) \url{https://doi.org/10.1016/S0081-1947(08)60132-0}.
\bibitem{Esh75} J.D. Eshelby, The elastic energy-momentum tensor, Journal of Elasticity 5 (1975) 321-335.
\bibitem{Fager} E,D. Fagerholm, W.M.C Foulkes, Y. Gallero- Salas, F. Helmchen, K.J. Friston, R.J. Moran, et al., Conservation laws by virtue of scale symmetries in neural systems. PLoS Comput Biol 16(5): e1007865, (2020). \url{https://doi.org/10.1371/journal. pcbi.1007865}
\bibitem{Fed} S. Federico, M.F. Alhasadi and A. Grillo, Eshelby's Inclusion Theory in the light of Noether's Theorem, Mathematics and Mechanics of Complex Systems, Vol. 7, No. 3, 2019, 247-285.
dx.doi.org/10.2140/memocs.2019.7.247
\bibitem{Frank} T. Frankel, The Geometry of Physics. An Introduction, Cambridge University Press, Cambridge UK, 2004.
\bibitem{Frisch} M. Frisch, Mechanism, principles, and Lorentz's cautious realism, Studies in History and Philosophy of Modern Physics 36 (2005) 659-679.
\bibitem{Giov} M. Giovanelli, Like Thermodynamics before Boltzmann. On the Emergence of Einstein's Distinction between Constructive and Principle Theories, \url{http://philsci-archive.pitt.edu/16829/1/Principles\%20-\%20FINAL.pdf}
\bibitem{Gunth} W. G\"{u}nther, \"{U}ber einige Randintegrale der Elastomechanik,  Abhandlungen der Braunschweiger Wissenschaftlichen Gesellschaft 14 (1962), 53-72.
\bibitem{HTH} J. Hanc, S. Tuleja and M. Hancova, Symmetries and conservation laws: Consequences of Noether's theorem, American Journal of Physics, 72, 428-435 (2004); doi: 10.1119/1.1591764
\bibitem{Havas} P. Havas, The Connection between Conservation Laws and Invariance Groups: Folklore, Fiction and Fact, Acta Phyica Austriaca 38 (1973) 145-167.
\bibitem{Heras} J.A. Heras, Electromagnetism in Euclidean four space: A discussion between God and the Devil,
American Journal of Physics 62(10), (1994) 914-916.
\bibitem{Hill} E.L. Hill, Hamilton's principle and the conservation theorems of mathematical physics, Rev. Mod. Phys. 23 (1951), 253-260.
\bibitem{Ibrag} N.H. Ibragimov, Integrating factors, adjoint equations and Lagrangians, Journal of Mathematical Analysis and Applications 318 (2006), 742-757.
\bibitem{Ib} N.H. Ibragimov, The answer to the question put to me by L.V. Ovsyannikov 33 years ago, Archives of ALGA, vol. 3 (2006) 53-80. ALGA Publications ISSN 1652-4934. \url{https://www.researchgate.net/publication/311456950_Archives_of_ALGA_vol_3}.
\bibitem{IK} N.H. Ibragimov and T. Kolsrud, Lagrangian Approach to Evolution Equations: Symmetries and Conservation Laws, Nonlinear Dynamics 36 (2004), 29-40.
\bibitem{Jackiw} R. Jackiw, Introducing scale symmetry, Physics Today, January 1972, 23-27.
\bibitem{Janssen} M. Janssen, `No Success Like Failure ...': Einstein's Quest for General Relativity, 1907-1920, in The Cambridge Companion to Einstein, M. Janssen and C. Lehner (eds.). Cambridge: Cambridge University Press, 2014; pp. 167-227.
\bibitem{KB} T. W. B. Kibble and F. H. Berkshire, \emph{Classical Mechanics}, Imperial College Press, London, 2004.
\bibitem{Knowles}  J.K. Knowles and Eli Sternberg, On a Class of Conservation Laws in Linearized and Finite Elastostatics, Arch. Rat. Mech. Anal. 44 (1972), 187-211.
\bibitem{K-S1} Y. Kosmann-Schwarzbach, The Noether Theorems, Invariance and Conservation Laws in the Twentieth Century, Sources and Studies in the History of Mathematics and Physical Sciences, Springer, 2010.
\bibitem{K-S2} Y. Kosmann-Schwarzbach, The Noether theorems in context, this volume, and \url{https://arxiv.org/pdf/2004.09254.pdf}. 
\bibitem{Blund} T. Lancaster and S.J. Blundell, Quantum Field Theory for the Gifted Amateur,  Oxford University Press, Oxford 2015.
\bibitem{Lange} M. Lange, Laws and meta-laws of nature: Conservation laws and symmetries. Studies in the History and
Philosophy of Modern Physics, 38, (2007) 457-481.
\bibitem{LL} L. D. Landau and E. M. Lifshitz,  \textit{Mechanics}, Pergamon Press, New York, 1960.
\bibitem{Lees} Lees, J. P., et al. (\emph{BABAR} Collaboration), 2012, Observation of time reversal violation in the $B^0$ meson system, Phys. Rev. Lett. 109, 211801.
\bibitem{LMA} L. Mart\'{i}nez Alonso, On the Noether map, Letters in Mathematical Physics 3, (1979) 419-424.
\bibitem{Noether} E. Noether, Invariante Variationsprobleme. G\"{o}ttinger Nachrichten, Mathematisch-physikalische Klasse 2 (1918), 235-257. 
\bibitem{Olver3} P.J. Olver, Conservation laws in elasticity I. General results, Arch. Rat. Mech. Anal. 85 (1984), 111-129. 
\bibitem{Olver4} P.J. Olver, Conservation laws in elasticity II. Linear homogeneous isotropic elastostatics, Arch Rat. Mech. Anal. 85 (1984), 131-160.
\bibitem{Olver86} P.J. Olver, Noether's theorems and systems of Cauchy-Kovalevskaya type, in \emph{Nonlinear Systems of Partial Differential
Equations in Applied Mathematics}, Lectures in Applied Mathematics Vol. 23 (American Mathematical Society, Providence,
R.I, 1986), pp. 81-104.
\bibitem{Olver} P.J. Olver, Applications of Lie Groups to Differential Equations, (New York: Springer-Verlag), 1986.
\bibitem{Olver18} P.J. Olver, Emmy Noether's enduring legacy in symmetry, Symmetry: Culture and Science, 29(4), 475-485 (2018).
\bibitem{Olver20} P.J. Olver, Divergence Invariant Variational Problems, 2020. \url{http://www-users.math.umn.edu/~olver/s_/divp.pdf}
\bibitem{Orz} C. A. Orzalesi, Charges and Generators of Symmetry Transformations in Quantum Field Theory, Reviews of Modern Physics 42(4) (1970) 381-408.
\bibitem{Pauli21} W. Pauli, Relativit\"{a}tstheorie, Encyklop\"{a}die der mathematischen Wissenschaften, Vol V19, B. G. Teubner, Liepsig, 1921.
\bibitem{Pauli} W. Pauli, The Theory of Relativity, Pergamon Press, Toronto, 1958.
\bibitem{Pesk} M.E. Peskin and D.V. Schroeder, An Introduction to Quantum Field Theory, Perseus Books, Cambridge, Mass, 1995.
\bibitem{RBL} J. Read, H.R. Brown and D. Lehmkuhl, Two Miracles in the General Theory of Relativity, Studies in History and Philosophy of Modern Physics 64, 14-25 (2018); \url{http://philsci-archive.pitt.edu/14589}.
\bibitem{Rice} J. R. Rice, Conserved Integrals and Energetic Forces, in Fundamentals of Deformation and Fracture (Eshelby Memorial Symposium), editors B.A. Bilby, K.J. Miller and J.R.  Willis,  (Cambridge: Cambridge University Press) 1985, pp, 43-56.
\bibitem{Rosen} J. Rosen, Redundancy and superfluity for electromagnetic fields and potentials, American Journal of Physics, 48, 1071-1073 (1980); doi: 10.1119/1.12289.
\bibitem{Ryder} L.H. Ryder, Quantum Field Theory, Cambridge University Press, 2nd edition, 1996. 
\bibitem{Sak} J.J. Sakurai, Invariance Principles and Elementary Particles, Princeton University Press, Princeton, 1964.
\bibitem{Sant} R.M. Santilli, Foundations of Theoretical Mechanics II, Birkhoffian Generalisation of Hamiltonian Mechanics, Springer-Verlag, New York, 1983.
\bibitem{Sch} D. Schroeren, The metaphysics of invariance, Studies in History and Philosophy of Modern Physics 70 (2020) 51-64.
\bibitem{Smith} S.R. Smith, Symmetries and the explanation of conservation laws in the light of the inverse problem
in Lagrangian mechanics, Studies in History and Philosophy of Modern Physics 39 (2008) 325-345.
\bibitem{SW} S. Steinberg and K.B. Wolf, Symmetry, Conserved Quantities and Moments in Diffusive Equations, J. Math. Anal. and Appl.80 (1981), 36-45.
\bibitem{Steudel62} H. Steudel, \"{U}ber die Zuordnung zwischen Invarianzeigenschaften und Erhaltungss\"{a}tzen, Z. Naturforscg. 17a (1962) 120-132.
\bibitem{Steudel} H. Steudel, Die Struktur der Invarianzgruppe fur lineare Feldtheorien, Zeit. Naturforsch. 21A (1966), 1826-1828.
\bibitem{Sud} A. Sudbery, A vector Lagrangian for the electromagnetic field,  J. Phys. A: Math. Gen. 19 (1986) L33-L36.
\bibitem{Sutton} A. P. Sutton, Physics of Elasticity and Crystal Defects, Oxford University Press, 2020.
ISBN: 9780198860785.
\bibitem{TB}  L.J. Tassie and H.A. Buchdahl, Gauge-independent theory of Symmetry. I,, Australian Journal of Physics, 17 (1964) 431-439.
\bibitem{Tavel} Emmy Noether (1971) Invariant variation problems, Transport Theory and Statistical Physics, 1:3, 186-207, DOI: 10.1080/00411457108231446
\bibitem{vF} B. C. van Fraassen, \emph{Laws and Symmetry}, Clarendon Press, Oxford, 1989.
\bibitem{Vino} A.M. Vinogradov, Local Symmetries and Conservation Laws, Acta Applicandae Mathematicae 2, (1984) 21-78.
\bibitem{Wein} S. Weinberg, Dreams of a final theory, Pantheon Press, New York, 1993. 
\bibitem{Wigner} E. Wigner, On the unitary representations of the inhomogeneous Lorentz group, Annals of Mathematics, 40 (1939) 149-204.
\bibitem{Wigner54} E. Wigner, Conservation Laws in Classical and Quantum Physics, Progress of Theoretical Physics, 11(4~5), 437-440.
\bibitem{Will} C.M. Will, The Confrontation between General Relativity and Experiment, \url{https://arxiv.org/pdf/1403.7377.pdf}.
\bibitem{Wolf} T. Wolf, A comparison of four approaches to the calculation of conservation laws,  European Journal of Applied Mathematics (2002), vol. 13, pp. 129-152. DOI 10.1017/S0956792501004715.
\bibitem{YM} C.N. Yang and R.L Mills, Conservation of Isotopic Spin and Isotopic Gauge Invariance, Phys Rev 96(1) (1954) 191-195.
\bibitem{Zee1} A. Zee, Fearful Symmetry: The Search for Beauty in Modern Physics, Princeton University Press 1999.
\bibitem{Zee2} A. Zee, Quantum Field Theory in a Nutshell, Princeton University Press, Princeton, 2003.

\end{thebibliography}

\section{Acknowledgments}
I am grateful to James Read, Bryan Roberts and Nicholas Teh for the invitation to contribute to this volume. Thanks go to Adam Caulton, Erik Curiel, and James Read for helpful comments, and to Niels Linnemann for spotting a technical error and making other suggestions for improvements. I am indebted to Adrian Sutton for illuminating discussion of Eshelby's contributions to the theory of elasticity. My greatest debt is to Peter Olver, for a generous and enlightening correspondence about various technical Noether-related matters.

\section{Appendices}

\appendix

\section{Heat equation I}

Consider two conserved quantities associated with the heat equation (17):
\begin{equation}
Q(u) = \int^{\infty}_{- \infty} u(x, t)\mbox{d}x,
\end{equation}
and
\begin{equation}
\tilde{Q}(u) = \int^{\infty}_{- \infty} xu(x, t)\mbox{d}x,
\end{equation}
which hold under the boundary conditions $u_x|^{\infty}_{- \infty} = 0$ and $(xu_x - u)|^{\infty}_{- \infty} = 0$, respectively\footnote{See \cite{SW}, p. 39.}, where $u_{x} \equiv \frac{\partial u}{\partial x}$. These conserved quantities cannot be associated in the Noetherian sense with symmetries of (17) because the heat equation does not have a variational formulation. However, consider the Lagrangian containing an auxiliary variable $v = v(x, t)$:
\begin{equation}
\mathcal{L} = \frac{1}{2} (vu_t -uv_t) + u_xv_x.
\end{equation}
The Euler-Lagrange equations obtained by varying with respect to $v$ and $u$ are respectively (17) and its adjoint:
\begin{equation}
v_t + v_{xx} = 0.
\end{equation}
Noether's first theorem can now be shown to lead to the conservation of $Q$ and $\tilde{Q}$ in relation to the respective symmetries of the Lagrangian (22):
\begin{equation}
x' = x , \; \; t' = t , \; \;v' = v +  \epsilon; \;   \;   \;  x' = x , \; \; t' = t , \; \;v'= v + x\epsilon'.\footnote{The application of Noether's theorem for  deriving the conservation of (22) is given in \cite{IK} on p. 35, putting $\phi = 0$. The corresponding result for (23) follows from the equations (3) - (9) on p. 30, \emph{op. cit}. A novel non-Noetherian route to (22) is found in \cite{Ib}, pp. 69, 70, using the Lagrangian $\mathcal{L} = v(u_t -u_{xx})$ and the Galilean symmetry of the heat equation. In all these cases the derivation of a conservation equation of the form $\mbox{d}C_1/ \mbox{d}t + \mbox{d}C_2/ \mbox{d}x = 0$ must be followed by integration over space and appeal to Gauss' divergence theorem.}
\end{equation}

Now the Schr\"{o}dinger equation  in quantum mechanics for the free particle in one spatial dimension,
\begin{equation}
\psi_t - \frac{i\hbar}{2m}\psi_{xx} = 0,
\end{equation}
is the complex analogue of the heat equation (17) with finite (complex) diffusivity. The Lagrangian can be written as
\begin{equation}
\mathcal{L} = \frac{i\hbar}{2}\left(\psi^*\psi_t - \psi^*_t \psi \right) - \frac{\hbar^2}{2m} \psi^*_x\psi_x,
\end{equation}
in analogy to (24), where $\psi^{\ast}$ is the complex conjugate of $\psi$.  It seems that the first Noether-type demonstration that strict wave-functional analogues exist of the conservation principles associated with the quantities (22) and (23), from analogues of the transformations (26), was published in 2004; the authors were unaware of the analogy with the heat equation.\footnote{See \cite{BH2}, section III.} Note that because the wave function is complex, so that $\psi^*$ is not an auxiliary function independent of $\psi$, the Schr\"{o}dinger equation is time-reversal invariant, unlike the heat equation (17).

\section{Heat equation II}

In relation to the heat equation (17), consider the general infinitesimal transformations
\begin{equation}
x' = x + \xi_x[u]\epsilon, \; \; t' = t + \xi_t[u]\epsilon, \; \;u'= u + \phi[u]\epsilon.
\end{equation}
In terms of  their generators, the combination of these transformations can be represented by a vector field over the space of dependent and independent variables:
\begin{equation}
\mathbf{v} = \xi_x[u]\partial_x + \xi_t[u]\partial_t  + \phi[u]\partial_u, 
\end{equation}
where $\partial_x \equiv \frac{\partial}{\partial x}$, etc. Its evolutionary form is\footnote{See  \cite{Olver}, equation (5.7).}
\begin{equation}
\tilde{\mathbf{v}} = \left[ \phi[u] - \left(\xi_x[u]\partial_x + \xi_t[u]\partial_t\right)\right]\partial_u. 
\end{equation}
The specific symmetries (18), (19) and (20) of the heat equation are represented by 
\begin{equation}
\mathbf{v}_{18} = \partial_t; \; \; \mathbf{v}_{19} = - \dot{u}\partial_u; \; \; \mathbf{v}_{20} = u_{xx}\partial_u,
\end{equation}
where $u_{xx} \equiv \frac{\partial^2u}{\partial x^2}$. It is easily seen that (i) $\mathbf{v}_{19}$ is the evolutionary form of $\mathbf{v}_{18}$, (ii) the evolutionary forms of both $\mathbf{v}_{18} -\mathbf{v}_{20}$ and $\mathbf{v}_{19} -\mathbf{v}_{20}$ vanish on-shell,  and (iii) the evolutionary form of $\mathbf{v}_{18} -\mathbf{v}_{19}$ is identically zero.

\section{Free particle}

\begin{enumerate}
\item In relation to the action defined with respect to the standard action $\mathcal{L} = \frac{1}{2} \dot{q}^2$ for the free particle, the nature of the variational symmetries in section 4.2 is as follows:

\begin{itemize} 

\item Boost transformations (7) are a quasi-symmetry, with divergence term $\dot{q}\epsilon$, and the equivalent (8) are a strict symmetry. The corresponding non-trivial conserved quantity in both cases is $q - \dot{q}t$.
 
 \item Time translation transformations (10) are a strict symmetry; (11) are a quasi-symmetry with divergence $\frac{\mbox{d}}{\mbox{d}t}\left(-\frac{q\dot{q}}{t}\right)\epsilon$; (12) are a quasi-symmetry with divergence $\frac{\mbox{d}}{\mbox{d}t}\left(- \frac{\dot{q}^2}{2}\right)\epsilon$ and (13) are a quasi-symmetry with divergence $\frac{\mbox{d}}{\mbox{d}t}\left( \frac{q\dot{q}}{t} -\frac{\dot{q}^2}{2}\right)\epsilon$. In all these equivalent cases the nontrivial conserved quantity is $\frac{\dot{q}^2}{2}$.
 
 \end{itemize}
 
 \item In relation to the action defined with respect to the Noether Lagrangian (9):
 
 \begin{itemize}
 
 \item Transformations (7) are a strict symmetry and transformations (8) are a quasi-symmetry with divergence $\frac{\mbox{d}}{\mbox{d}t}\left( -\frac{2q}{\dot{q}^2} \mathcal{L}\right)\epsilon$, where $\mathcal{L}$ is the Lagrangian (9). The conserved quantity as above.
 
 \item Time translation transformations (10) are a quasi-symmetry with divergence $ \frac{\mbox{d}}{\mbox{d}t}\left(\frac{q^2}{2t^2}\right)\epsilon$; (11) are a quasi-symmetry with divergence $\frac{\mbox{d}}{\mbox{d}t}\left(\frac{3q^2}{2t^2} - \frac{q\dot{q}}{t} -\frac{q^3}{\dot{q}t^3}\right)\epsilon$; (12) are a quasi-symmetry with divergence $\frac{\mbox{d}}{\mbox{d}t}\left( \frac{q\dot{q}}{t} -\frac{\dot{q}^2}{2}\right)\epsilon$ and (13) are a strict symmetry. The conserved quantity as above.
 
 \end{itemize}
 
 \item Consider now the group of rescaling transformations defined by 
 \begin{equation}
q' = \lambda q; \: \: t' = \lambda^2 t, \;  \;  \;  \;  \; \lambda \in  \mathbb{R} \setminus \{0\}.
\end{equation}
Both of the Lagrangians above are strictly invariant with respect to these transformations, and the associated conserved quantity is $\dot{q}(q -\dot{q}t)$, the product of the conserved quantities associated with spatial translations and boosts, respectively.\footnote{For recent discussions, see \cite{Olver20}, and \cite{Fager} where an application to neural systems is given.} When a potential energy term is included in the standard Lagrangian, the transformations (22) are variational symmetries only if the particle moves under the influence of an inverse cube force law. More interesting applications of (approximate) scale invariance have been studied in relativistic quantum field theory.\footnote{For an introduction see \cite{Jackiw}.}
 
\end{enumerate}

\end{document}